\documentclass[preprint,12pt]{elsarticle}

\usepackage{amssymb}
\usepackage{amsmath}
\usepackage{graphicx}
\usepackage{booktabs}
\usepackage{multirow}
\usepackage{threeparttable}
\usepackage{makecell}
\usepackage{hyperref}
\usepackage{url}
\usepackage{rotating}

\journal{Nuclear Physics B}

\begin{document}

\begin{frontmatter}



\title{Advancing Long-Tailed Pediatric Arrhythmia Classification with a Novel Contrastive Loss and Multimodal Learning}

\author[aff1]{Yiqiao Chen\fnref{fn1}}
\author[aff1]{Zijian Huang\fnref{fn1}}
\author[aff1]{Zhenghui Feng\corref{cor1}}

\cortext[cor1]{Corresponding author.}
\ead{fengzhenghui@hit.edu.cn}

\affiliation[aff1]{organization={Faculty of Frontier Sciences, Harbin Institute of Technology},
            city={Shenzhen},
            postcode={518055},
            country={China}}

\fntext[fn1]{Yiqiao Chen and Zijian Huang contributed equally to this work.}

\begin{abstract}
Arrhythmias are a major cause of sudden cardiac death in children, making automated rhythm classification from electrocardiograms (ECGs) clinically important. However, pediatric arrhythmia analysis remains challenging because of age-dependent waveform variability, limited data availability, and a pronounced long-tailed class distribution that hinders recognition of rare but clinically important rhythms. To address these issues, we propose a multimodal end-to-end framework that integrates surface ECG and intracardiac electrogram (IEGM) signals for pediatric arrhythmia classification. The model combines dual-branch feature encoders, attention-based cross-modal fusion, and a lightweight Transformer classifier to learn complementary electrophysiological representations. We further introduce an Adaptive Global Class-Aware Contrastive Loss (AGCACL), which incorporates prototype-based alignment, class-frequency reweighting, and globally informed hard-class modulation to improve intra-class compactness and inter-class separability under class imbalance. We evaluate the proposed method on the pediatric subset of the Leipzig Heart Center ECG-Database and establish a reproducible preprocessing pipeline including rhythm-segment construction, denoising, and label grouping. The proposed approach achieves 96.22\% Top-1 accuracy and improves macro precision, macro recall, macro F1 score, and macro F2 score by 4.48, 1.17, 6.98, and 7.34 percentage points, respectively, over the strongest baseline. These results indicate improved minority-sensitive classification performance on the current benchmark. However, further validation under subject-independent and multicenter settings is still required before clinical translation.
\end{abstract}

\begin{keyword}
Adaptive global class-aware contrastive Loss (AGCACL), multimodal learning, pediatric, arrhythmia classification, long-tailed
\end{keyword}
\end{frontmatter}

\section{Introduction}
\begin{figure*}[htbp]
	\centering
	\includegraphics[width=0.95\textwidth]{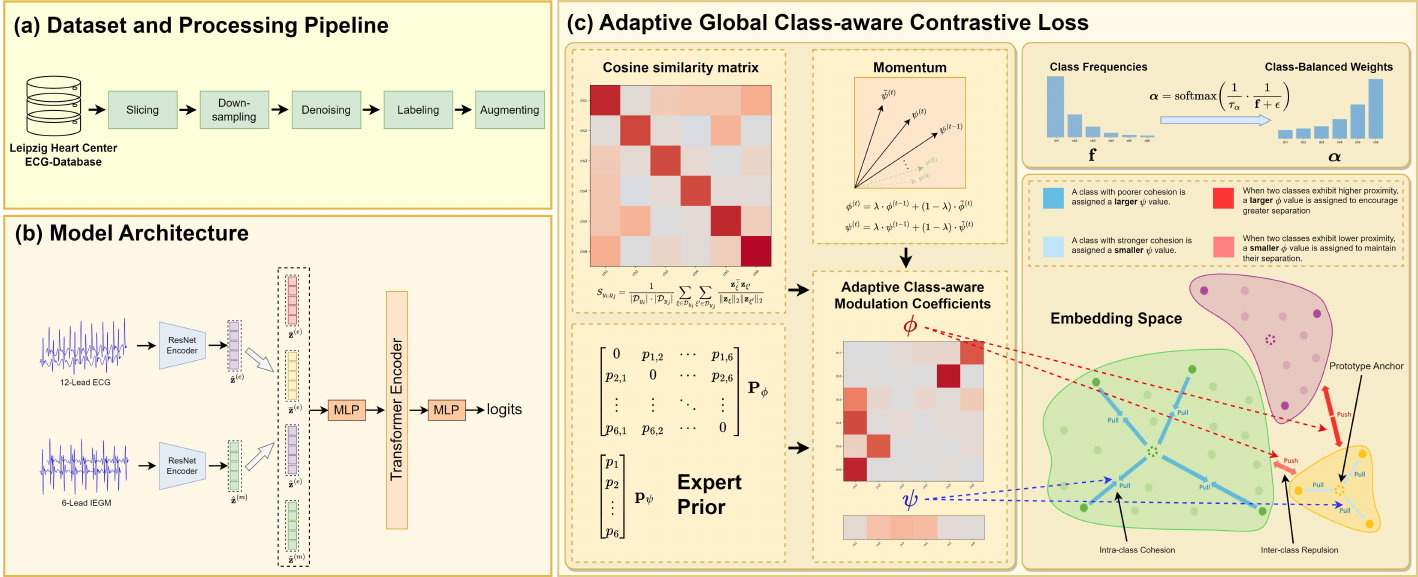}
	\caption{Overview of the proposed framework for long-tailed pediatric arrhythmia classification. The framework comprises dataset preprocessing, multimodal ECG–IEGM feature learning, and optimization with the proposed Adaptive Global Class-Aware Contrastive Loss (AGCACL).}
	\label{fig:framework}
\end{figure*}

\par According to the World Health Organization's 2025 report, cardiovascular diseases remain the leading cause of death globally~\cite{who2025cvd}. Among the various manifestations of cardiovascular disease, cardiac arrhythmias are of particular clinical concern due to their high prevalence and strong association with adverse outcomes such as heart failure and stroke~\cite{hu2022transformer,yang2024associations}. In addition to increasing morbidity and mortality, arrhythmias may compromise hemodynamic stability, impair quality of life, and create substantial demands on healthcare resources through repeated surveillance and intervention. Therefore, the development of reliable methods for arrhythmia detection and classification is of major importance for early diagnosis, clinical decision-making, and the prevention of serious cardiovascular events.

\par Artificial intelligence (AI) has shown great potential in electrocardiography (ECG) analysis and cardiovascular disease assessment, enabling more efficient and automated extraction of clinically relevant information from cardiac signals~\cite{li2025electrocardiogram}. In this context, AI-assisted arrhythmia classification has been extensively investigated as a promising approach for improving the efficiency and reliability of cardiac rhythm analysis. As the standard non-invasive tool for cardiac monitoring, ECG has consequently become the primary signal source in the majority of arrhythmia classification studies~\cite{liu2024adaptive, ebrahimi2020review}. Kim \textit{et al.} proposed an automatic arrhythmia classification framework that combines a residual network with squeeze-and-excitation blocks and bidirectional LSTM, and evaluated it on eight-, four-, and two-class settings using MIT-BIH Arrhythmia Database, MIT-BIH Atrial Fibrillation Database, and the PhysioNet/CinC 2017 database, respectively, showing superior overall and minority-class performance compared with conventional methods \cite{kim2022automatic}. Akan \textit{et al.} developed ECGformer, a Transformer-based model for heartbeat arrhythmia classification that splits ECG signals into patches and exploits self-attention to model long-range dependencies, and validated it on a preprocessed five-class dataset derived from MIT-BIH and PTB ECG records, where it achieved strong performance for automated arrhythmia recognition \cite{akan2023ecgformer}. Alamatsaz \textit{et al.} proposed a lightweight hybrid CNN--LSTM model for ECG-based arrhythmia detection, in which preprocessed ECG signals from the MIT-BIH Arrhythmia Database and the Long-Term AF Database were classified into eight arrhythmias and normal rhythm, achieving a mean diagnostic accuracy of 98.24\%, while SHAP was further used to improve model interpretability \cite{alamatsaz2024lightweight}. Venkatesh \textit{et al.} developed a 1D-CNN--BiLSTM ensemble for automated atrial arrhythmia classification and evaluated it on Lead-II ECG signals from the Chapman University and Shaoxing People’s Hospital dataset, where the model classified sinus rhythm, sinus tachycardia, atrial tachycardia, atrial fibrillation, and atrial flutter with an accuracy of 94\% \cite{venkatesh2024automated}. Guhdar \textit{et al.} proposed a hybrid 1D-CNN framework with a squeeze-and-excitation attention mechanism for automated ECG arrhythmia classification, and validated it on the MIT-BIH Arrhythmia Database and PTB Diagnostic ECG Database, achieving accuracies of 99.48\% and 99.83\%, respectively, and 99.64\% on the combined dataset, with corresponding F1-scores of 0.99, 1.00, and 1.00 \cite{guhdar2025advanced}.

\par Although ECG-based arrhythmia classification has been extensively investigated, especially through advanced computational methods such as artificial intelligence, research has predominantly centered on adult populations, but pediatric cases have been largely overlooked \cite{mayourian2024pediatric}, despite the vital clinical importance of timely and accurate diagnosis in children to prevent adverse outcomes such as syncope, heart failure, and sudden cardiac death. At the same time, it is clear that directly applying techniques developed for adult populations to pediatric cases is inherently biased, because pediatric arrhythmias differ significantly from adult arrhythmias in both clinical manifestations and ECG characteristics. For instance, supraventricular tachycardia occurs more commonly in children \cite{kafali2022common}, and certain rhythm patterns considered abnormal in adults, such as sinus arrhythmia or wandering pacemaker, can be normal in pediatric patients \cite{drago2018neonatal}. Additionally, compared to adults, pediatric ECGs typically show faster heart rates, shorter PR and QRS intervals, and age-dependent changes in waveform morphology \cite{bratincsak2020electrocardiogram}. 

\par Class imbalance refers to a learning scenario in which different classes are represented by markedly unequal numbers of samples, often causing predictive models to bias toward majority classes while underperforming on minority ones. This is a persistent challenge in arrhythmia classification, as publicly available datasets often contain substantially more samples from common rhythm categories than from rare abnormal classes \cite{xiao2023deep,ansari2023deep}. This issue also exists in pediatric arrhythmia classification and may become even more severe because some abnormal pediatric rhythms are inherently uncommon and difficult to acquire in sufficient quantity. When the skew is extreme, the resulting class distribution can further exhibit a long-tailed pattern, in which a small number of head classes account for most observations while tail classes contain only very limited data. Such a phenomenon is clearly observed in the six-class classification setting adopted in this study: the cumulative duration of the head class, Sinus rhythm, reaches 37,048 s, whereas that of the tail class, Tachycardias, is only 53 s, yielding an imbalance ratio of approximately 699:1.

\par These challenges suggest that improving pediatric arrhythmia classification requires not only more robust learning strategies for long-tailed data, but also richer electrophysiological representations than those provided by surface ECG alone. Although ECG remains the standard modality for rhythm assessment, it primarily reflects body-surface manifestations of cardiac electrical activity and may be insufficient to fully characterize complex pediatric arrhythmias. In contrast, intracardiac electrogram (IEGM) signals record electrical activity closer to the source of cardiac conduction and can therefore provide complementary information for rhythm discrimination. Motivated by this, we propose a multimodal deep learning framework that jointly exploits ECG and IEGM signals for long-tailed pediatric arrhythmia classification, while introducing an Adaptive Global Class-Aware Contrastive Loss (AGCACL) to improve representation learning under severe class imbalance. An overview of the proposed method is presented in Fig.~\ref{fig:framework}. Overall, the main contributions of this paper are as follows:
\begin{itemize}
	\item To the best of our knowledge, this is the first work to perform automated pediatric arrhythmia classification on the Leipzig Heart Center dataset, for which we establish a complete preprocessing pipeline, including signal slicing, denoising, and relabeling.
	\item We propose a novel multimodal deep learning framework that integrates CNN-based encoders, a dual-branch semantic attention fusion module, and a Transformer Encoder-based classification head to fully exploit temporal and cross-modal dependencies.
	\item To mitigate class imbalance and long-tailed challenges inherent in the dataset, we introduce a novel Adaptive Global Class-Aware Contrastive Loss (AGCACL), which incorporates class-wise reweighting, prototype-based alignment, and globally modulated inter-class repulsion.
	\item Extensive experiments demonstrate that the proposed multimodal framework and the Adaptive Global Class-Aware Contrastive Loss (AGCACL) consistently outperform competitive baseline methods across multiple evaluation metrics.
\end{itemize}

\par The rest of the paper is organized as follows: Section~\ref{Methods} describes the dataset processing pipeline, model architecture, proposed contrastive loss, training strategies, and experimental setup. Section~\ref{Results and Discussion} presents the results and discussion. Finally, Section~\ref{Conclusion} concludes the paper.

\section{Methods}\label{Methods}
\subsection{Dataset and Preprocessing Pipeline}
This study utilizes the Leipzig Heart Center ECG-Database~\cite{klehsleipzig}, which was publicly released on the PhysioNet platform in 2025 \cite{goldberger2000physiobank}. The database focuses on rare but clinically significant arrhythmias in pediatric and congenital heart disease (CHD) populations. The dataset contains recordings from 39 subjects, comprising a total of 1,075.85 minutes of 12-lead ECG and intracardiac electrograms (IEGM), with 113,924 annotated heartbeats. All annotations were manually performed by two ECG specialists. Signals were sampled at 977 Hz, with filter settings of 0.05–100 Hz. This study focuses specifically on the pediatric subset of the database, consisting of 29 children, all of whom were clinically diagnosed with arrhythmias. To the best of our knowledge, this is the first study to employ the Leipzig Pediatric Arrhythmia Database for deep learning–based rhythm classification, paving the way for intelligent ECG analysis tailored to pediatric patients.
\par
\begin{figure}[htbp]
	\centering
	\includegraphics[width=0.95\textwidth]{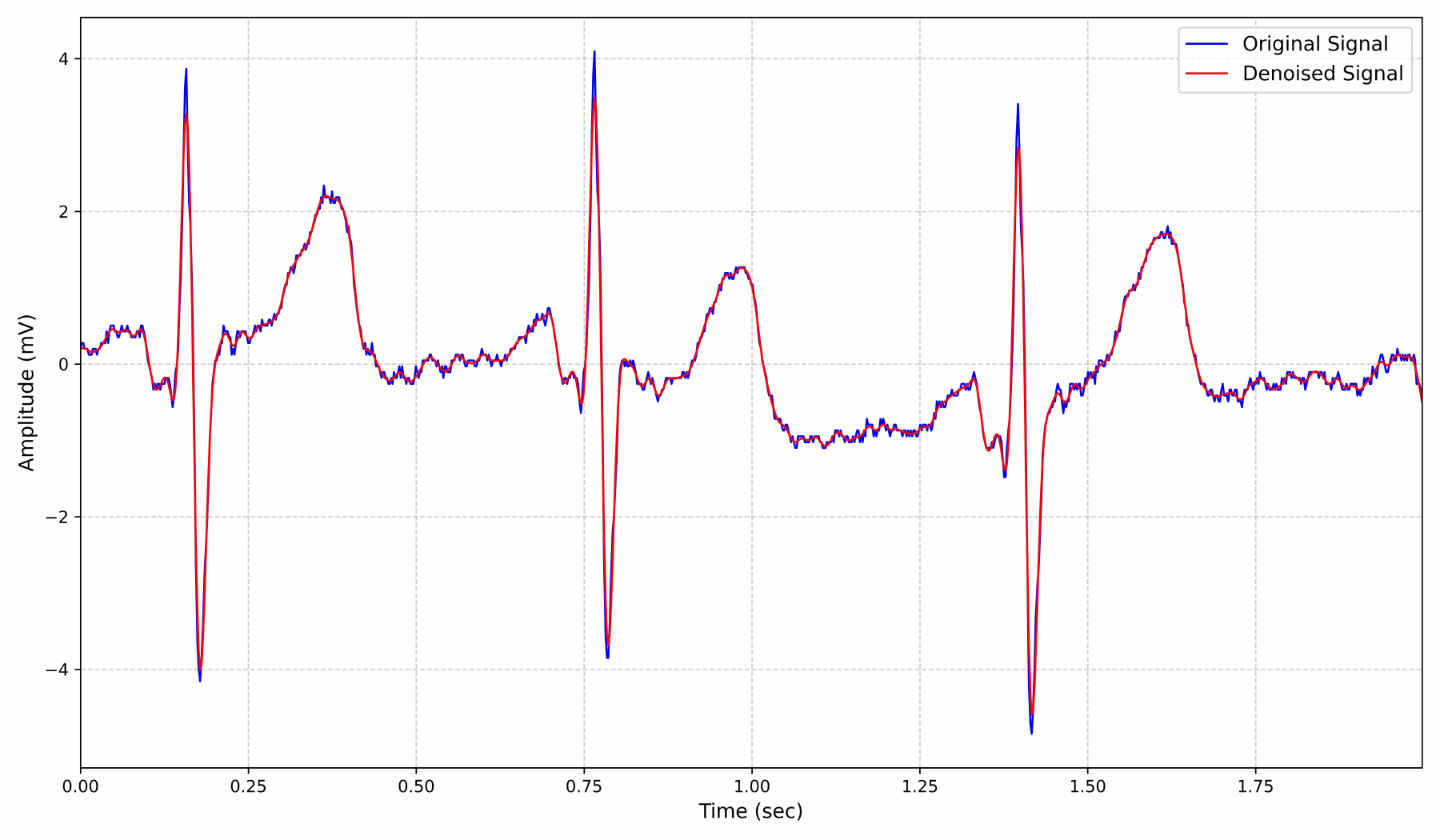}
	\caption{Representative single-lead ECG segment before and after denoising. The denoised waveform preserves the main morphological structure while reducing high-frequency noise.}
	\label{fig:denoise}
\end{figure}

\begin{figure*}[htbp]
	\centering
	\includegraphics[width=0.95\textwidth]{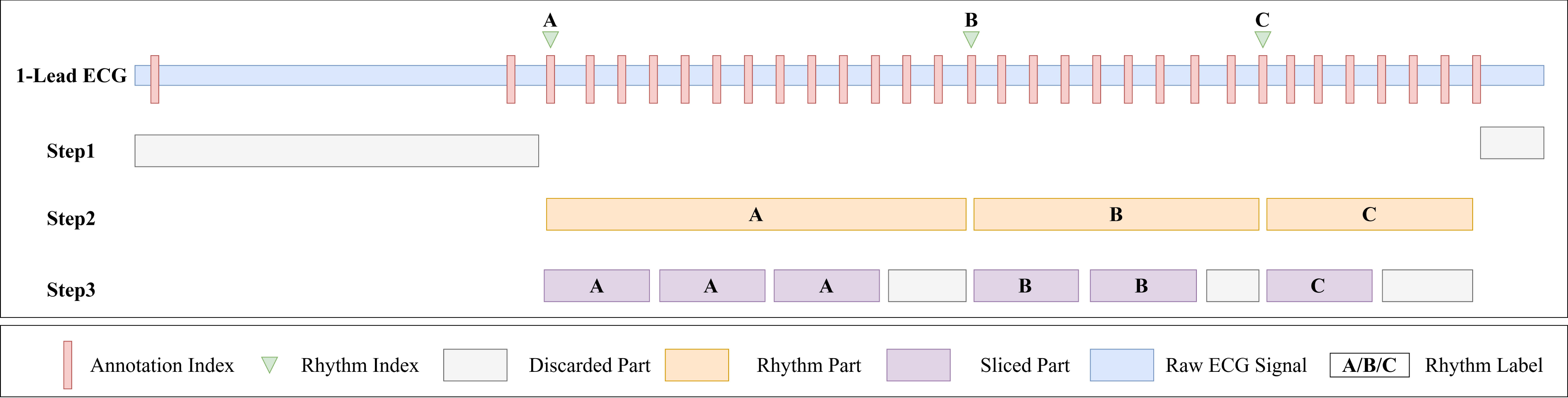}
	\caption{Illustration of the ECG preprocessing pipeline. Step~1 discards unannotated signal segments. Step~2 segments the remaining signal based on rhythm labels provided by experts. Step~3 slices each rhythm segment into fixed 2-second windows to ensure consistent labeling. Only windows fully contained within a single rhythm type are preserved. This pipeline ensures the generation of structurally consistent and label-pure ECG segments for model training.
	}
	\label{fig:slice}
\end{figure*}

 To ensure consistent lead configuration across all 29 subjects, we retained only the following 18 leads: [I, II, III, aVR, aVL, aVF, V1, V2, V3, V4, V5, V6, RVA12, CS12, CS34, CS56, CS78, CS90]. To enable standardized model training and evaluation, we built a complete data preprocessing pipeline from scratch. Our signal slicing and labeling process is illustrated in Fig.~\ref{fig:slice}. Specifically, we first selected 29 pediatric subjects (x001–x029) clinically diagnosed with arrhythmias from the database and retained only the signal segments annotated with rhythm labels. These segments were then divided according to their rhythm labels and further segmented into 2-second windows to ensure clear rhythm attribution for each sample. To reduce computational cost, all samples were uniformly downsampled by a factor of two. To further enhance signal quality, we sequentially applied z-score normalization, a 4th-order Chebyshev low-pass filter (cutoff frequency: 45Hz) to remove high-frequency noise, and finally soft-thresholding wavelet denoising \cite{donoho1994ideal} based on the db6 wavelet with decomposition level of 5 to mitigate physiological artifacts caused by spectral aliasing. The resulting dataset comprises rhythm segments that are structurally consistent, quality-controlled, and suitable for deep learning modeling. A comparison of signals before and after denoising is shown in Fig.~\ref{fig:denoise}.

\par In the aforementioned slicing and labeling process, each extracted segment was assigned a fine-grained low-level label corresponding to the Aux String annotations defined in the official WFDB files. For the arrhythmia classification task in this study, these low-level labels were further consolidated into high-level labels according to the label definitions provided by the original dataset authors. To ensure consistency between label statistics and our rhythm-segment-based sample construction pipeline, we additionally summarized the pediatric subset at the episode level rather than the beat level. Specifically, consecutive rhythm annotations in the WFDB files were used to define rhythm episodes, from which the episode count and cumulative duration of each low-level rhythm label were computed and then aggregated into their corresponding high-level categories, as summarized in Table~\ref{tab:rhythm_grouping}. As shown in Table~\ref{tab:rhythm_grouping}, the pediatric subset exhibits a pronounced long-tailed distribution at the episode level: sinus rhythm, supraventricular tachycardia, and paced rhythms account for the vast majority of total annotated duration, whereas categories such as atrial tachycardia, ectopic rhythm, and tachycardias are represented by only a small number of episodes and much shorter cumulative durations. This long-tailed label distribution further motivates the use of high-level labels as the classification targets in all experiments. This design choice was motivated by two factors. First, some low-level labels are absent or severely underrepresented in the pediatric subset. Specifically, (IVR) is not observed, and labels such as (AFL) occur only rarely, making reliable partitioning across training, validation, and test sets difficult. Second, grouping low-level labels into clinically meaningful high-level categories improves interpretability and supports better generalization across datasets and real-world clinical scenarios. Based on this grouping strategy, the dataset was partitioned into training, validation, and test subsets in a 7:1:2 ratio, while preserving the overall high-level class distribution across splits as much as possible. Owing to the extremely limited number of pediatric subjects and the sparsity of several rhythm categories, the dataset was partitioned at the window level rather than by strictly subject-independent splitting.
\begin{table}[htbp]
	\centering
	\renewcommand{\arraystretch}{1.18}
	\caption{Episode-level distribution of low-level rhythm annotations and grouped high-level classes in the pediatric subset of the Leipzig Heart Center ECG-Database.}
	\label{tab:rhythm_grouping}
	\begin{threeparttable}
	\begin{tabular}{c l l c c c}
		\toprule
		\makecell[c]{\textbf{Class}\\\textbf{ID}} 
		& \makecell[c]{\textbf{High-level}\\\textbf{Label}} 
		& \makecell[c]{\textbf{Low-level}\\\textbf{Label}} 
		& \makecell[c]{\textbf{Episode}\\\textbf{Count}} 
		& \makecell[c]{\textbf{Duration}\\\textbf{(s)}} 
		& \makecell[c]{\textbf{Class Total}\\\textbf{(s)}} \\
		\midrule
		
		1 & Sinus rhythm & (N & 727 & 37048 & 37048 \\
		
		\addlinespace[1.4ex]
		\multirow{2}{*}{2}
		& \multirow{2}{*}{\makecell[l]{Supraventricular\\[-0.2ex]Tachycardia}}
		& (AVRT  & 93  & 4381 & \multirow{2}{*}{7076} \\
		& & (AVNRT & 126 & 2695 & \\
		
		\addlinespace[1.4ex]
		\multirow{2}{*}{3}
		& \multirow{2}{*}{\makecell[l]{Paced\\[-0.2ex]rhythms}}
		& (/A & 454 & 5068 & \multirow{2}{*}{7030} \\
		& & (/V & 196 & 1961 & \\
		
		\addlinespace[1.4ex]
		\multirow{3}{*}{4}
		& \multirow{3}{*}{\makecell[l]{Atrial\\[-0.2ex]Tachycardia}}
		& (AFIB & 12 & 490 & \multirow{3}{*}{542} \\
		& & (EAT  & 4  & 15  & \\
		& & (AFL  & 1  & 37  & \\
		
		\addlinespace[1.4ex]
		\multirow{3}{*}{5}
		& \multirow{3}{*}{\makecell[l]{Ectopic\\[-0.2ex]rhythm}}
		& (A & 8  & 171 & \multirow{3}{*}{352} \\
		& & (B & 2  & 11  & \\
		& & (J & 25 & 170 & \\
		
		\addlinespace[1.4ex]
		\multirow{2}{*}{6}
		& \multirow{2}{*}{Tachycardias}
		& (VT  & 12 & 53 & \multirow{2}{*}{53} \\
		& & (IVR & 0  & 0  & \\
		
		\bottomrule
	\end{tabular}
	\begin{tablenotes}[flushleft]
		\footnotesize
		\item Note: Class ID denotes the numeric label index used in this study, ranging from 1 to 6. High-level Label denotes the grouped diagnostic category used in this study, and Low-level Label denotes the original rhythm annotation in the WFDB files. Episode Count indicates the number of rhythm episodes for each low-level label. Duration (s) denotes the cumulative duration of that low-level label in seconds, and Class Total (s) denotes the summed duration of all low-level labels within the same high-level category.
	\end{tablenotes}
	\end{threeparttable}
\end{table}
\par In addition, to address the class imbalance inherent in the original dataset, we employed data augmentation strategies to upsample minority classes to a target size of 3000 samples per class. Specifically, we applied a range of physiologically plausible signal transformations—including temporal shifting, polarity flipping, time warping, baseline drifting and amplitude scaling—to generate synthetic ECG samples. All augmentations were applied channel-wise with controlled intensity. This augmentation was applied exclusively to the training set, while the validation and test sets remained unchanged. 
\par Finally, each augmented sample was separated into two distinct synchronized signal modalities: 12-lead ECG and 6-lead IEGM, which were processed independently in subsequent multimodal learning stages.

\subsection{Proposed Multimodal Architecture}
To overcome the limited feature learning capacity of single-modality ECG models and to exploit the complementary physiological information embedded in IEGM signals, we propose a multimodal end-to-end model that jointly processes both modalities for collaborative feature learning. The model is designed to capture fine-grained morphological patterns from ECG and IEGM signals while achieving semantic alignment and information complementarity through a cross-modal semantic fusion module. 
Specifically, given the two modalities for each sample: the standard 12-lead surface ECG signal $\mathbf{X}^{(e)} \in \mathbb{R}^{B \times 12 \times 977}$ and a 6-lead IEGM signal $\mathbf{X}^{(m)} \in \mathbb{R}^{B \times 6 \times 977}$, the network outputs unnormalized classification logits $\mathbf{y} \in \mathbb{R}^{B \times C}$, where $B$ is the batch size and $C=6$ is the number of diagnostic classes (High-level label). During inference, a softmax is applied to the logits to obtain class probabilities, and the class with the highest probability is selected as the final prediction.
\subsubsection{ResNet-based Modality Encoding Process}
First, each modality is processed independently by a modified 1D ResNet encoder \cite{he2016deep} to extract fine-grained morphological feature vectors:
\begin{align}
	\hat{\mathbf{Z}}^{(e)} &= \mathrm{ResNet}_e(\mathbf{X}^{(e)}) \in \mathbb{R}^{B \times 256}\\
	\hat{\mathbf{Z}}^{(m)} &= \mathrm{ResNet}_m(\mathbf{X}^{(m)}) \in \mathbb{R}^{B \times 256}
\end{align}
The encoder begins with a $7{\times}1$ convolution (stride 2), batch normalization, and a non-inplace ReLU, followed by a $3{\times}1$ max pooling (stride 2). Next, Four residual stages progressively increase the channel dimension $\{64,128,256,256\}$ while reducing temporal resolution, with two residual blocks per stage. Then, Each block adopts a pre-activation-like pattern: \emph{Conv–BN–ReLU, then Conv–BN, followed by identity addition and a final ReLU}, with an optional $1{\times}1$ projection on the skip path when dimensions differ. Last, adaptive average pooling reduces the temporal dimension to one, producing a fixed 256-D vector per branch.

\subsubsection{Attention-based Cross-modal Semantic Alignment Fusion}
Let $\mathbf{W}^{(q_\cdot)}$, $\mathbf{W}^{(k_\cdot)}$, and $\mathbf{W}^{(v_\cdot)} \in \mathbb{R}^{256\times 256}$ and
$\mathbf{b}^{(q_\cdot)}, \mathbf{b}^{(k_\cdot)}, \mathbf{b}^{(v_\cdot)} \in \mathbb{R}^{256}$ be learnable projection weights and biases for queries, keys, and values, respectively. Given the encoder outputs $\hat{\mathbf{Z}}^{(e)}, \hat{\mathbf{Z}}^{(m)} \in \mathbb{R}^{B\times 256}$, the gated cross-modal features are computed as
\begin{align}
	\bar{\mathbf{Z}}^{(e)} &=
	\sigma\!\left(
	\frac{
		\big(\hat{\mathbf{Z}}^{(e)}\mathbf{W}^{(q_e)}+\mathbf{b}^{(q_e)}\big)
		\odot
		\big(\hat{\mathbf{Z}}^{(m)}\mathbf{W}^{(k_m)}+\mathbf{b}^{(k_m)}\big)
	}{\sqrt{h}}
	\right)
	\notag\\
	&\quad \odot
	\big(\hat{\mathbf{Z}}^{(m)}\mathbf{W}^{(v_m)}+\mathbf{b}^{(v_m)}\big)
	\\[0.3em]
	\bar{\mathbf{Z}}^{(m)} &=
	\sigma\!\left(
	\frac{
		\big(\hat{\mathbf{Z}}^{(m)}\mathbf{W}^{(q_m)}+\mathbf{b}^{(q_m)}\big)
		\odot
		\big(\hat{\mathbf{Z}}^{(e)}\mathbf{W}^{(k_e)}+\mathbf{b}^{(k_e)}\big)
	}{\sqrt{h}}
	\right)
	\notag\\
	&\quad \odot
	\big(\hat{\mathbf{Z}}^{(e)}\mathbf{W}^{(v_e)}+\mathbf{b}^{(v_e)}\big)
\end{align}
where $\odot$ denotes the Hadamard (elementwise) product, $h{=}256$, and $\sigma(\cdot)$ is the sigmoid function. For each sample in the batch, the gated and original feature vectors are concatenated along the feature dimension and then fused through a feedforward layer with residual normalization:
\begin{equation}
	\mathbf{z} = \mathrm{LN}\!\Big(\mathrm{ReLU}\big([\bar{\mathbf{Z}}^{(e)},\bar{\mathbf{Z}}^{(m)},\hat{\mathbf{Z}}^{(e)},\hat{\mathbf{Z}}^{(m)}]\mathbf{W}^f+\mathbf{b}^f\big)\Big)
\end{equation}
where $\mathbf{W}^{(f)} \in \mathbb{R}^{1024\times 512}$, $\mathbf{b}^{(f)} \in \mathbb{R}^{512}$, $\mathbf{z} \in \mathbb{R}^{B\times 512}$ is the fused feature vector, and $\mathrm{LN}$ is the LayerNorm function. We apply dropout rate as $p{=}0.1$ inside the fusion layer before layer normalization.

\subsubsection{Transformer-based Classification Head}

To further capture dependencies among the dimensions of the fused representation, we employ a lightweight Transformer-based classification head on top of the fused cross-modal feature. Although the fused representation is not a temporal sequence in the conventional sense, we treat feature dimensions as ordered tokens to enable interaction modeling across latent attributes. The positional encoding here serves as an index-aware parameterization rather than a physiological time prior. Let the fused feature be denoted by $\mathbf{Z} \in \mathbb{R}^{B \times 512}$, where $B$ is the batch size. Rather than directly feeding $\mathbf{Z}$ into a multilayer perceptron, we interpret each 512-dimensional fused vector as a scalar sequence of length $L{=}512$. Concretely, an additional singleton dimension is appended so that the input is represented as a sequence in $\mathbb{R}^{B \times L \times 1}$.

Each scalar token is then projected into a higher-dimensional embedding space through a learnable linear projection:
\begin{equation}
	\mathbf{U} = \mathbf{Z}' \mathbf{W}^p + \mathbf{b}^p,
\end{equation}
where $\mathbf{Z}' \in \mathbb{R}^{B \times L \times 1}$ denotes the expanded sequence representation of $\mathbf{Z}$, $\mathbf{W}^p \in \mathbb{R}^{1 \times d}$ and $\mathbf{b}^p \in \mathbb{R}^{d}$ are the projection parameters, and $\mathbf{U} \in \mathbb{R}^{B \times L \times d}$ is the token embedding sequence. In our implementation, the embedding dimension is set to $d{=}64$.

To preserve the order information of the sequence, we add a standard sinusoidal positional encoding \cite{vaswani2017attention} to $\mathbf{U}$, yielding the Transformer input:
\begin{equation}
	\mathbf{H}_0 = \mathbf{U} + \mathrm{PE}.
\end{equation}
The resulting sequence $\mathbf{H}_0$ is then processed by a lightweight Transformer encoder consisting of a single encoder layer. The encoder uses multi-head self-attention with $h{=}4$ heads, a feed-forward hidden dimension of $d_{\mathrm{ff}}{=}128$, dropout rate $p{=}0.1$, GELU activation, and a pre-normalization design for improved training stability. The encoded representation is given by
\begin{equation}
	\mathbf{H}_1 = \mathrm{Encoder}(\mathbf{H}_0; d, h, d_{\mathrm{ff}}, p),
\end{equation}
where $\mathbf{H}_1 \in \mathbb{R}^{B \times L \times d}$.

To obtain a global representation for classification, we apply average pooling over the sequence dimension:
\begin{equation}
	\bar{\mathbf{h}} = \frac{1}{L} \sum_{t=1}^{L} \mathbf{H}_1(:, t, :) ,
\end{equation}
where $\bar{\mathbf{h}} \in \mathbb{R}^{B \times d}$ denotes the pooled feature representation. This pooled feature is subsequently passed through layer normalization and dropout, and finally mapped to the class space by a fully connected classifier:
\begin{equation}
	\mathbf{y} = \mathrm{Dropout}(\mathrm{LayerNorm}(\bar{\mathbf{h}})) \mathbf{W}^c + \mathbf{b}^c,
\end{equation}
where $\mathbf{W}^c \in \mathbb{R}^{d \times C}$ and $\mathbf{b}^c \in \mathbb{R}^{C}$ are the classifier parameters, and $\mathbf{y} \in \mathbb{R}^{B \times C}$ denotes the output logits. In our setting, $C{=}6$ is the number of diagnostic classes.

This design allows the classifier head to model interactions among feature dimensions in a lightweight yet expressive manner. Compared with a plain linear classifier, the Transformer encoder can capture contextual dependencies within the fused representation, while the use of a single encoder layer keeps the computational overhead low and remains suitable for stable end-to-end training.

\subsection{Adaptive Global Class-aware Contrastive Loss}
Contrastive learning has been widely applied in biomedical representation learning under supervised~\cite{shiri2024supervised}, self-supervised~\cite{azizi2021big}, and semi-supervised~\cite{hu2021semi} paradigms. Here, we focus on the supervised paradigm~\cite{khosla2020supervised}. However, the traditional supervised contrastive loss relies on pairwise intra-class attraction, but for underrepresented classes, this can lead to unstable optimization and noisy gradients, resulting in scattered or fragmented embeddings. Moreover, traditional supervised contrastive loss functions usually treat all sample pairs equally which uniformly pull intra-class pairs together and push inter-class pairs apart. While effective in balanced settings, traditional supervised contrastive loss functions become suboptimal under imbalance circumstance, where minority classes receive insufficient optimization focus.
\par The instability of pairwise intra-class alignment, especially in long-tailed scenarios, has been mitigated by introducing statistical or learnable class prototypes to provide stable semantic anchors and enhance class-wise cohesion~\cite{alonso2021semi,cui2021parametric}. Meanwhile, to address limitations of symmetric contrastive objectives under class imbalance, \cite{zhu2022balanced} introduces an averaging-based correction to balance the influence across classes, but it does not account for sample difficulty. Similarly, RCL also applies class-number-based weighting to mitigate imbalance, but still overlooks the mining of hard samples \cite{de2023long}. In contrast, many hard negative mining methods adopt instance-pair-level weighting based solely on similarity~\cite{yuan2025weighted, hou2023improving, srinivasa2023cwcl, li2022targeted}, lacking global class-level statistics and exhibiting high sensitivity to batch composition—issues that become more severe under long-tailed settings.
\par To address the above limitations, we propose a novel loss function called Adaptive Global Class-aware Contrastive Loss (AGCACL). We retain the previous notations, where $\mathbf{z}_i$ denotes the embedding of sample $x_i$, $y_i$ its corresponding label, $N$ the batch size, the total loss is defined as:
\begin{align}
	\mathcal{L}_{\text{AGCACL}} &= \frac{1}{N} \sum_{i=1}^{N} \alpha_{y_i} \left(\ell_{\text{intra}}^{(i)} + \ell_{\text{inter}}^{(i)} \right) \label{eq:agcacl} \\
	\ell_{\text{intra}}^{(i)} &= \psi_{y_i} \cdot \left(1 - \frac{\mathbf{z}_i^\top \mathbf{c}_{y_i}}{\|\mathbf{z}_i\|_2 \|\mathbf{c}_{y_i}\|_2} \right) \label{eq:intra_instance} \\
	\ell_{\text{inter}}^{(i)} &= \log \sum\limits_{\substack{j=1 \\ j \ne i}}^{N} \phi_{y_i, y_j} \cdot \exp\left( \frac{\mathbf{z}_i^\top \mathbf{z}_j}{\tau \|\mathbf{z}_i\|_2 \|\mathbf{z}_j\|_2} \right) \label{eq:inter_instance} 
\end{align}
Here, $\phi_{y_i, y_j}$ denotes a class-wise repulsion coefficient that modulates the inter-class similarity contribution, and $\psi_{y_i}$ controls the strength of intra-class alignment for each class. These parameters allow the model to adaptively emphasize certain class relationships during optimization. $\mathbf{c}_{y_i}$ denotes the learnable prototype of class $y_i$, which is jointly optimized during training to represent the semantic center of the corresponding class in the embedding space. $\alpha_{y_i}$ adjusts the overall loss contribution of class $y_i$ to account for data imbalance. $\alpha_{y_i}$ is a class-specific balancing weight designed to mitigate the influence of class imbalance by reweighting each instance's contribution to the total loss. Finally, $\tau$ is a temperature parameter that scales the inter-class contrastive component $\ell_{\text{inter}}$ to control the softness of similarity scores. In the following, we describe the rationale and computation of each key component in AGCACL. We begin with the intra-class alignment term and class-level weighting, followed by the design of the inter-class contrastive formulation. Finally, we present our adaptive class-aware modulation coefficients, which serve as a core contribution of this framework.
\par As shown in Equation \eqref{eq:intra_instance}, to improve representation stability in long-tailed regimes, we replace pairwise intra-class attraction with a center-based alignment mechanism, where a learnable class prototype $\mathbf{c}_{y_i}$ serves as the semantic anchor for each category. This formulation is inspired by the parametric contrastive framework introduced in \cite{cui2021parametric}, which incorporates class-wise learnable centers into contrastive learning. By treating class prototypes as jointly optimized parameters, this approach enables more consistent gradients and facilitates learning under limited data conditions.
However, unlike intra-class attraction, inter-class separation aims to model local decision boundaries. Therefore, we retain the pairwise formulation to reflect finer class-level distinctions within the batch.
\par To reduce the dominance of majority classes, we assign each instance a class-aware weight $\alpha_{y_i}$ derived from the inverse class frequency over the \emph{entire} training set (excluding data augmentations). Let $\mathbf{f}$ be the vector of class frequencies, where each entry corresponds to the number of samples in a particular class, and let $\epsilon>0$ be a small constant. We compute a temperature-scaled softmax over inverse frequencies:
\begin{equation}
	\alpha = \mathrm{softmax}\!\left(\frac{1}{\tau_{\alpha}}\cdot\frac{1}{\mathbf{f}+\epsilon}\right)
\end{equation}
and use $\alpha_{y_i}$ to reweight the loss of sample $i$. Here, $\mathbf{f}$ is estimated from the full training set.
\par Beyond static weighting, our key contribution lies in the introduction of adaptive class-aware modulation terms, $\phi_{y_i, y_j}$ and $\psi_{y_i}$, which are collectively referred to as global modulation coefficients. These terms respectively control inter-class repulsion and intra-class alignment strength. To compute these class-aware modulation coefficients, we begin by constructing a global class-to-class similarity matrix $\mathbf{S} \in \mathbb{R}^{C \times C}$. Here, we distinguish between instance indices $i, j$, which refer to samples within a batch, and class labels $y_i, y_j \in \{1, \dots, C\}$, which represent the corresponding class identities. Each element $S_{y_i, y_j}$ represents the average pairwise cosine similarity between all training embeddings of class $y_i$ and class $y_j$:
\begin{equation}
	S_{y_i, y_j} = \frac{1}{|\mathcal{D}_{y_i}| \cdot |\mathcal{D}_{y_j}|} \sum_{\xi \in \mathcal{D}_{y_i}} \sum_{\xi' \in \mathcal{D}_{y_j}} \frac{\mathbf{z}_\xi^\top \mathbf{z}_{\xi'}}{\|\mathbf{z}_\xi\|_2 \|\mathbf{z}_{\xi'}\|_2}
	\label{eq:class_similarity}
\end{equation}
Here, $\mathcal{D}_{y_i}$ and $\mathcal{D}_{y_j}$ denote the index sets of all training samples belonging to class $y_i$ and $y_j$, respectively. That is, $\xi$ and $\xi'$ represent sample indices such that $\mathbf{z}_\xi$ and $\mathbf{z}_{\xi'}$ are their corresponding embeddings.
This similarity matrix serves as a globally informed semantic map, allowing the model to reason over full-dataset statistics rather than relying on potentially noisy or sparse batch-level interactions. Such global perspective is particularly beneficial in long-tailed settings, where batch samples may not adequately reflect class-level distributions.
Next, to derive the intra-class alignment weight $\psi_{y_i}$, we focus on the diagonal elements $S_{y_i, y_i}$, which reflect the internal cohesiveness of each class $y_i$. Classes with low self-similarity (i.e., high internal variance) require stronger alignment forces. Thus, we take the inverse of $S_{y_i, y_i}$ and apply a temperature-scaled softmax normalization to obtain the intra-class modulation weights:
\begin{equation}
	\psi_{y_i} = \frac{\exp\left( \frac{1}{S_{y_i,y_i} + \epsilon} / \tau_{\psi} \right)}{\sum\limits_{m=1}^{C} \exp\left( \frac{1}{S_{m,m} + \epsilon} / \tau_{\psi} \right)}
	\label{eq:psi}
\end{equation}
where $\epsilon$ is a small constant for stability, and $\tau_{\psi}$ controls the temperature. This formulation is both class-aware and semantically grounded, as it dynamically adjusts each class's alignment strength based on its empirical compactness, while also allowing prior knowledge injection if needed.
In contrast, for inter-class repulsion weighting $\phi_{y_i, y_j}$, we emphasize relative separation among distinct classes. For each anchor class $y_i$, we compute softmax attention over the off-diagonal elements of $S_{y_i, n}$:
\begin{equation}
	\phi_{y_i, y_j} =
	\begin{cases}
		\frac{\exp\left(S_{y_i, y_j} / \tau_{\phi}\right)}{\sum\limits_{\substack{n=1 \\ n \ne y_i}}\limits^{C} \exp\left(S_{y_i, n} / \tau_{\phi}\right)} & \text{if } y_i \ne y_j \\
		0 & \text{if } y_i = y_j
	\end{cases}
	\label{eq:phi}
\end{equation}
where $\tau_{\phi}$ is a temperature coefficient. This design allows the model to selectively emphasize harder negatives (i.e., classes more similar to the anchor), enabling a more nuanced and structurally interpretable inter-class contrastive signal. By framing contrastive repulsion in terms of soft attention over class-level similarity, we ensure that the model allocates training focus according to the semantic difficulty of each class pair, rather than treating all inter-class interactions uniformly as in conventional pairwise schemes.
\par To further incorporate domain-specific semantic priors, we extend the formulation of $\phi_{y_i, y_j}$ and $\psi_{y_i}$ by integrating user-defined prior knowledge. Specifically, we define a class-to-class semantic prior matrix $\mathbf{P}_\phi \in \mathbb{R}^{C \times C}$ and an intra-class semantic prior vector $\mathbf{p}_\psi \in \mathbb{R}^{C}$, where each element encodes the strength of prior similarity or cohesion desired between specific classes. Let $\bar{\phi}$ and $\bar{\psi}$ denote the average values of the computed $\phi_{y_i, y_j}$ and $\psi_{y_i}$ across classes. We then refine the original coefficients as follows:
\begin{align}
	\tilde{\phi}_{y_i, y_j} &= \phi_{y_i, y_j} + \mathbf{P}_\phi[y_i, y_j] \cdot \bar{\phi} \\
	\tilde{\psi}_{y_i} &= \psi_{y_i} + \mathbf{p}_\psi[y_i] \cdot \bar{\psi}
\end{align}
By default, the prior matrices $\mathbf{P}_\phi$ and $\mathbf{p}_\psi$ are initialized to zero, making this augmentation optional. When specified, they allow explicit control over the relative strength of class interactions based on expert knowledge.
\par Meanwhile, to stabilize training and prevent oscillations due to frequent updates, we adopt a momentum-based update scheme for both $\phi$ and $\psi$. At each epoch $t$, the coefficients are updated as exponential moving averages of their current and previous values:
\begin{align}
	\phi^{(t)} &= \lambda \cdot \phi^{(t-1)} + (1 - \lambda) \cdot \tilde{\phi}^{(t)} \\
	\psi^{(t)} &= \lambda \cdot \psi^{(t-1)} + (1 - \lambda) \cdot \tilde{\psi}^{(t)}
\end{align}
Here, $\tilde{\phi}^{(t)}$ and $\tilde{\psi}^{(t)}$ denote the raw coefficients computed at epoch $t$ (including semantic prior adjustment), and $\lambda \in [0,1)$ is the momentum factor controlling the influence of historical estimates. At the first epoch ($t=0$), $\phi^{(0)}$ and $\psi^{(0)}$ are initialized as the directly computed $\tilde{\phi}^{(0)}$ and $\tilde{\psi}^{(0)}$, respectively.
\par Overall, our formulation provides an adaptive, semantically grounded, and class-aware framework for regulating contrastive dynamics across intra- and inter-class relations. In particular, the AGCACL possesses the following properties: (1) \textbf{Globally informed}: Both $\phi_{y_i, y_j}$ and $\psi_{y_i}$ are derived from the full training set, rather than individual batches, which is especially important in long-tailed scenarios where batch composition may be highly variable. (2) \textbf{Semantically grounded}: The formulation allows the incorporation of semantic priors or domain-specific class relations, enabling optional manual adjustments to inject expert knowledge. (3) \textbf{Data-driven and adaptive}: Coefficients are updated with a momentum-based smoothing strategy, enabling stable adaptation to evolving embeddings while maintaining compatibility with stochastic training and random initialization. (4) \textbf{Class-aware and robust}: Unlike uniform pair-wise weighting, our class-level coefficients $\phi$ and $\psi$ adaptively modulate alignment and repulsion per class, improving resilience to imbalance and semantic ambiguity, especially in long-tailed settings. (5) \textbf{Simple and interpretable}: The framework avoids complex optimization by relying on straightforward similarity measures and structured normalization for inter-class modeling. Together, these properties enable $\phi$ and $\psi$ to evolve smoothly with the embedding space, ensuring training stability and effective integration of optional prior knowledge under stochastic optimization and random initialization.
\subsection{Training Strategies}
In this study, the training objective is composed of Focal Loss and the proposed Adaptive Global Class-Aware Contrastive Loss (AGCACL), which are combined with equal weight. Focal Loss with a focusing parameter $\gamma=1.0$ is employed to downweight easy samples and emphasize hard ones, while AGCACL leverages class prototypes and a global similarity matrix to enhance intra-class compactness and inter-class separability. During training, a custom class-balanced sampler was used to maintain an approximately uniform class distribution within each mini-batch. The optimizer is Adam with a learning rate and weight decay of 1e-4, batch size of 48, and a total of 30 epochs, without learning rate scheduling, gradient clipping, or mixed precision. For AGCACL, the temperature parameters are set as $\tau=0.1$, $\tau_{\phi}=0.01$, $\tau_{\psi}=0.1$, and $\tau_{\alpha}=0.1$, with a momentum coefficient of 0.9. Class prototypes are initialized randomly in a 512-dimensional space and jointly optimized with network parameters. Class-frequency priors are computed from the empirical class distribution of the original training set, excluding augmented samples. In addition, a semantic prior matrix is defined, where only the entries at positions $(6,3)$, $(3,6)$, $(5,1)$, and $(1,5)$ are assigned a value of 1, while all other entries remain 0, thereby enforcing stronger repulsion between these selected pairs of easily confused classes. The semantic prior matrix was defined based on domain knowledge and tuned using the validation set only, without any access to or optimization on the test set. This training strategy enables more stable optimization under long-tailed settings, improving both robustness and clinical relevance.

\subsection{Experimental Setup}
All experiments were conducted on the Leipzig Heart Center pediatric arrhythmia dataset. Data augmentation was applied only to the training set, while validation and test sets remained unchanged. For baselines, we reproduced the original architectures with minor adaptations (e.g., six-class output, consistent use of lead~II, removal of irrelevant branches), and all models were trained from scratch without external pretraining. To facilitate controlled comparison under a unified reproduction protocol, Focal Loss with $\gamma=1.0$ was uniformly adopted, and hyperparameters were aligned or lightly tuned on the validation set. For the comparison of contrastive losses, we additionally modified Focal InfoNCE into a supervised version by treating samples from the same class as positives. Experiments were conducted using two AMD EPYC 7B12 CPUs (64 cores, 128 threads, 2.25\,GHz base frequency) and eight NVIDIA RTX~4090 GPUs (24\,GB memory per GPU, 16{,}384 CUDA cores). Model performance was evaluated by Top-1 Accuracy, Macro Specificity, Precision, Recall, F1, and F2, where macro-averaging ensures equal weight to minority classes under class imbalance. Note that the proposed model leverages multimodal ECG–IEGM inputs, whereas reproduced baselines were adapted to the unified single-lead setting whenever their original architectures were not directly compatible with the present data format.

\section{Experimental Results and Discussion}\label{Results and Discussion}
\subsection{Comparison with State-of-the-Art (SOTA) Reproduced Baselines}
\begin{sidewaystable}[htbp]
	\centering
	\caption{Performance comparison of the proposed method and reproduced state-of-the-art baselines on the pediatric arrhythmia classification task. Results are reported in terms of Top-1 accuracy and macro-averaged specificity, precision, recall, F1, and F2.}
	\renewcommand{\arraystretch}{1.3}
	\begin{tabular}{lcccccc}
		\toprule
		Methods & Top-1 Acc & Macro S & Macro P & Macro R & Macro F1 & Macro F2 \\
		\midrule
		Akan et al. \cite{akan2023ecgformer} & 65.64 & 92.70 & 34.39 & 57.71 & 34.39 & 37.97 \\
		Alamatsaz et al. \cite{alamatsaz2024lightweight} & 90.69 & 97.36 & 62.17 & 76.14 & 65.95 & 70.38 \\
		Kim et al. \cite{kim2022automatic} & 95.07 & 98.55 & 85.68 & 85.19 & 84.26 & 84.54 \\
		Guhdar et al. \cite{guhdar2025advanced} & 86.38 & 96.87 & 60.61 & 76.86 & 61.11 & 65.42 \\
		Liu et al. \cite{liu2024adaptive} & 89.37 & 97.03 & 58.45 & 81.32 & 63.39 & 70.06 \\
		Mayourian et al. \cite{mayourian2024pediatric} & 93.98 & 98.50 & 75.61 & 91.40 & 79.74 & 84.67 \\
		Venkatesh et al. \cite{venkatesh2024automated} & 93.72 & 97.98 & 74.40 & 79.61 & 76.43 & 78.16 \\
		\textbf{Proposed} & $\mathbf{96.22}$ & $\mathbf{98.82}$ & $\mathbf{90.16}$ & $\mathbf{92.57}$ & $\mathbf{91.24}$ & $\mathbf{92.01}$ \\
		\bottomrule
	\end{tabular}
	\label{tab:comparison}
\end{sidewaystable}

\begin{table}[htbp]
	\centering
	\caption{Relative and absolute improvements of our proposed method 
		over the strongest baseline value for each metric. $\Delta$: absolute gain; \%$\uparrow$: relative gain.}
	\renewcommand{\arraystretch}{1.2}
	\setlength{\tabcolsep}{6pt}
	\begin{tabular}{lcccc}
		\toprule
		{Metric} & {Best Prior} & {Proposed} & {$\Delta$} & {\%$\uparrow$} \\
		\midrule
		Top-1 Acc (\%)   & 95.07 & 96.22 & +1.15 & +1.21 \\
		Macro S (\%)     & 98.55 & 98.82 & +0.27 & +0.27 \\
		Macro P (\%)     & 85.68 & 90.16 & +4.48 & +5.23 \\
		Macro R (\%)     & 91.40 & 92.57 & +1.17 & +1.28 \\
		Macro F1 (\%)    & 84.26 & 91.24 & +6.98 & +8.28 \\
		Macro F2 (\%)    & 84.67 & 92.01 & +7.34 & +8.67 \\
		\bottomrule
	\end{tabular}
	\label{tab:improve}
\end{table}

\begin{figure}[htbp]
	\centering
	\includegraphics[width=0.95\textwidth]{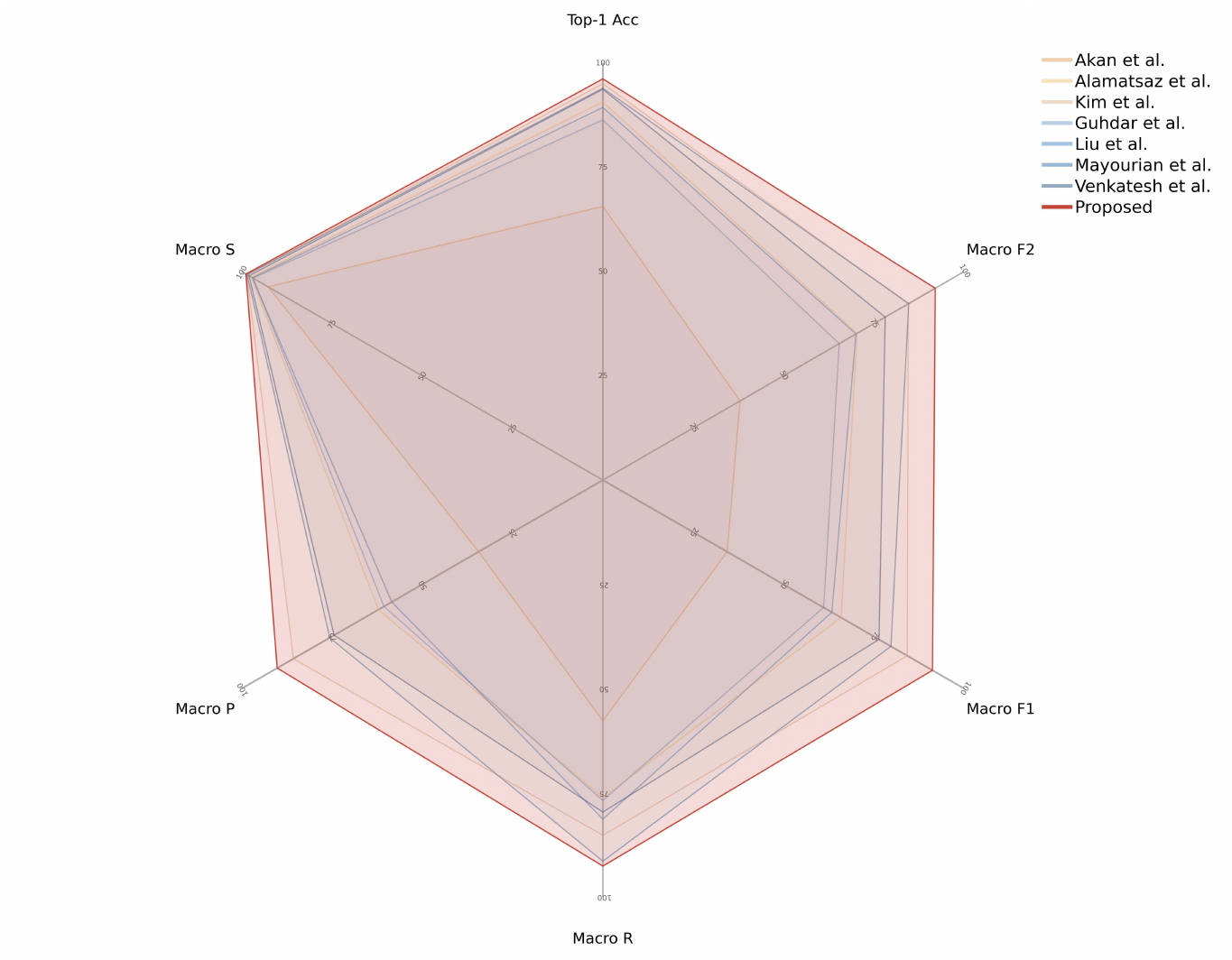}
	\caption{Radar-chart comparison of the proposed method and reproduced state-of-the-art baselines across six evaluation metrics.}
	\label{fig:radar}
\end{figure}

\begin{figure}[htbp]
	\centering
	\includegraphics[width=0.95\textwidth]{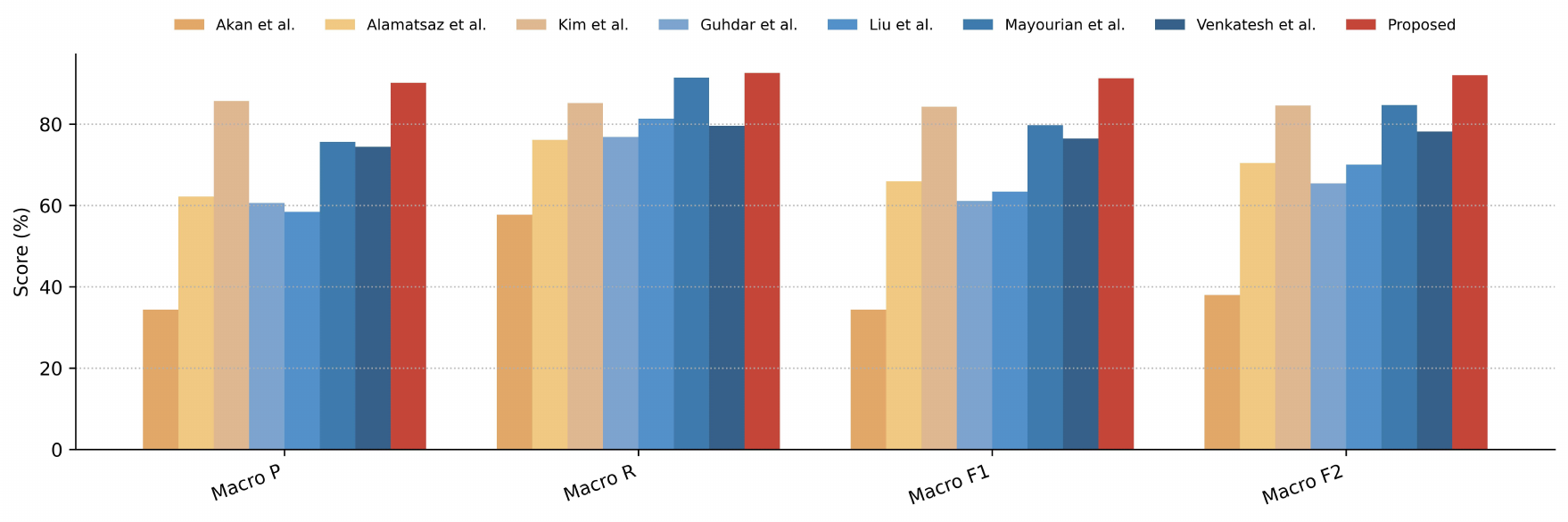}
	\caption{Bar plot illustrating detailed performance comparisons on four 
		macro-averaged metrics: Macro Precision, Macro Recall, Macro F1, and Macro F2.}
	\label{fig:bar}
\end{figure}

\par To comprehensively evaluate the effectiveness of the proposed framework, we compared it against seven representative state-of-the-art arrhythmia classification methods under a unified experimental setting. These baselines cover diverse architectural paradigms, including convolution-based models, lightweight deep networks, and Transformer-based approaches, thereby providing a broad reference for performance assessment. For fairness, all reproduced methods were adapted to the same pediatric arrhythmia classification task, trained from scratch on the same data split, and evaluated using the same metric set. The comparison therefore aims not only to assess whether the proposed method improves overall classification accuracy, but more importantly to examine whether it yields better balanced performance under the long-tailed pediatric setting, especially on macro-averaged metrics that are more sensitive to minority-class recognition. The quantitative results are summarized in Table~\ref{tab:comparison} and Table~\ref{tab:improve}, while Fig.~\ref{fig:radar} and Fig.~\ref{fig:bar} provide complementary visual comparisons across methods.

\par As shown in Table~\ref{tab:comparison} and Fig.~\ref{fig:radar}, the proposed method achieves the best overall performance among all compared baselines, reaching 96.22\% Top-1 Accuracy, 98.82\% Macro Specificity, 90.16\% Macro Precision, 92.57\% Macro Recall, 91.24\% Macro F1, and 92.01\% Macro F2. Notably, our method is the only one that consistently ranks first across all six evaluation metrics, indicating not only strong overall discriminative ability but also a more balanced performance across classes. Among the reproduced baselines, the strongest prior competitors are Kim et al.~\cite{kim2022automatic} and Mayourian et al.~\cite{mayourian2024pediatric}; however, our method still surpasses them on every reported metric. These results show that the proposed framework achieves the strongest overall performance among the reproduced baselines under the present experimental setting.

\par A closer inspection of Table~\ref{tab:improve} and Fig.~\ref{fig:bar} further shows that the performance gains are not uniformly distributed across metrics. In particular, the most pronounced improvements are observed on Macro Precision (+4.48 percentage points), Macro F1 (+6.98 percentage points), and Macro F2 (+7.34 percentage points), whereas the gains in Top-1 Accuracy and Macro Specificity are relatively modest. This pattern is important. It suggests that the proposed method does not merely improve predictions on already dominant or easy classes, but more importantly enhances the balance between false positives and false negatives across the full label space. Since Macro F1 and especially Macro F2 are more sensitive to class-imbalanced recognition quality, these improvements indicate stronger robustness under the long-tailed pediatric arrhythmia setting and suggest improved detectability of underrepresented and clinically important rhythms.

\par The differences among baseline methods are also consistent with the characteristics of this task. In particular, the Transformer-based ECGFormer of Akan et al.~\cite{akan2023ecgformer} yields the weakest overall performance. A likely reason is that pure Transformer-style architectures usually benefit from large-scale training data and long-sequence dependency modeling, whereas the present setting is substantially different: all reproduced baselines were adapted to a unified single-lead input protocol, the available pediatric data are limited, and each sample corresponds to a short 2-second rhythm segment. Under such conditions, the advantage of global self-attention is less fully exploited, while the model may become more sensitive to data scarcity and optimization instability. In contrast, methods with stronger convolutional inductive bias, such as Kim et al.~\cite{kim2022automatic}, achieve more competitive results because local morphological patterns remain highly informative in short ECG segments. Similarly, pediatric-related deep learning models such as Mayourian et al.~\cite{mayourian2024pediatric} show stronger recall-oriented performance, likely because their architectures are better aligned with clinically heterogeneous pediatric signals. Nevertheless, despite these differences, the baseline methods still exhibit limited macro-level performance, implying that existing approaches do not sufficiently address the long-tailed label distribution and the inadequate coverage of minority classes in this dataset.

\par The superiority of our method is likely attributable to the complementary effects of its four key components. First, the dual-branch ResNet encoders provide modality-specific representation learning for ECG and IEGM, enabling the network to capture both surface-level and intracardiac electrophysiological patterns. Second, the cross-modal semantic attention fusion module promotes feature alignment and selectively emphasizes mutually informative representations across the two modalities, thereby improving multimodal integration beyond simple concatenation. Third, the lightweight Transformer head models global dependencies within the fused representation while avoiding the heavy data demand of large Transformer backbones. Finally, the proposed AGCACL explicitly enhances intra-class compactness and inter-class separability under severe imbalance by combining prototype-based alignment, class-aware reweighting, and globally informed hard-class modulation. Taken together, these design choices yield a more discriminative and more balanced representation space, which is reflected by the consistent improvements in macro-averaged metrics.

\par From a clinical perspective, these findings are also meaningful. In pediatric arrhythmia analysis, errors on rare but clinically important rhythms may carry greater consequences than errors on majority classes. Therefore, the observed improvements in macro-averaged recall, F1, and F2 suggest that the proposed method may help reduce the risk of missed detection for uncommon arrhythmias while maintaining strong overall accuracy. These findings suggest potential utility for minority-sensitive rhythm recognition, although further subject-independent and multicenter validation is required before clinical deployment.

\subsection{Effectiveness of AGCACL: Comparison with Contrastive Losses}
\begin{sidewaystable}[htbp]
	\centering
	\caption{Performance comparison of different contrastive loss under a unified experimental setting. Results are reported using Top-1 accuracy and macro-averaged specificity, precision, recall, F1, and F2.}
	\renewcommand{\arraystretch}{1.3}
	\begin{tabular}{lcccccc}
		\bottomrule
		Methods & Top-1 Acc & Macro S & Macro P & Macro R & Macro F1 & Macro F2 \\
		\midrule
		None & \textbf{96.34} & 98.58 & 88.10 & 88.27 & 87.54 & 87.83 \\
		SupCon \cite{khosla2020supervised} & 96.30 & 98.58 & 83.48 & 85.65 & 84.34 & 85.06 \\
		RCL \cite{de2023long} & 95.05 & 97.79 & \textbf{93.04} & 82.64 & 87.14 & 84.31 \\
		Focal InfoNCE \cite{hou2023improving} & 95.62 & 98.74 & 83.65 & 91.78 & 86.73 & 89.43 \\
		\textbf{AGCACL} & 96.22 & \textbf{98.82} & 90.16 & \textbf{92.57} & \textbf{91.24} & \textbf{92.01} \\
		\bottomrule
	\end{tabular}
	\label{tab:loss_comparison}
\end{sidewaystable}

\begin{figure}[htbp]
	\centering
	\includegraphics[width=0.95\textwidth]{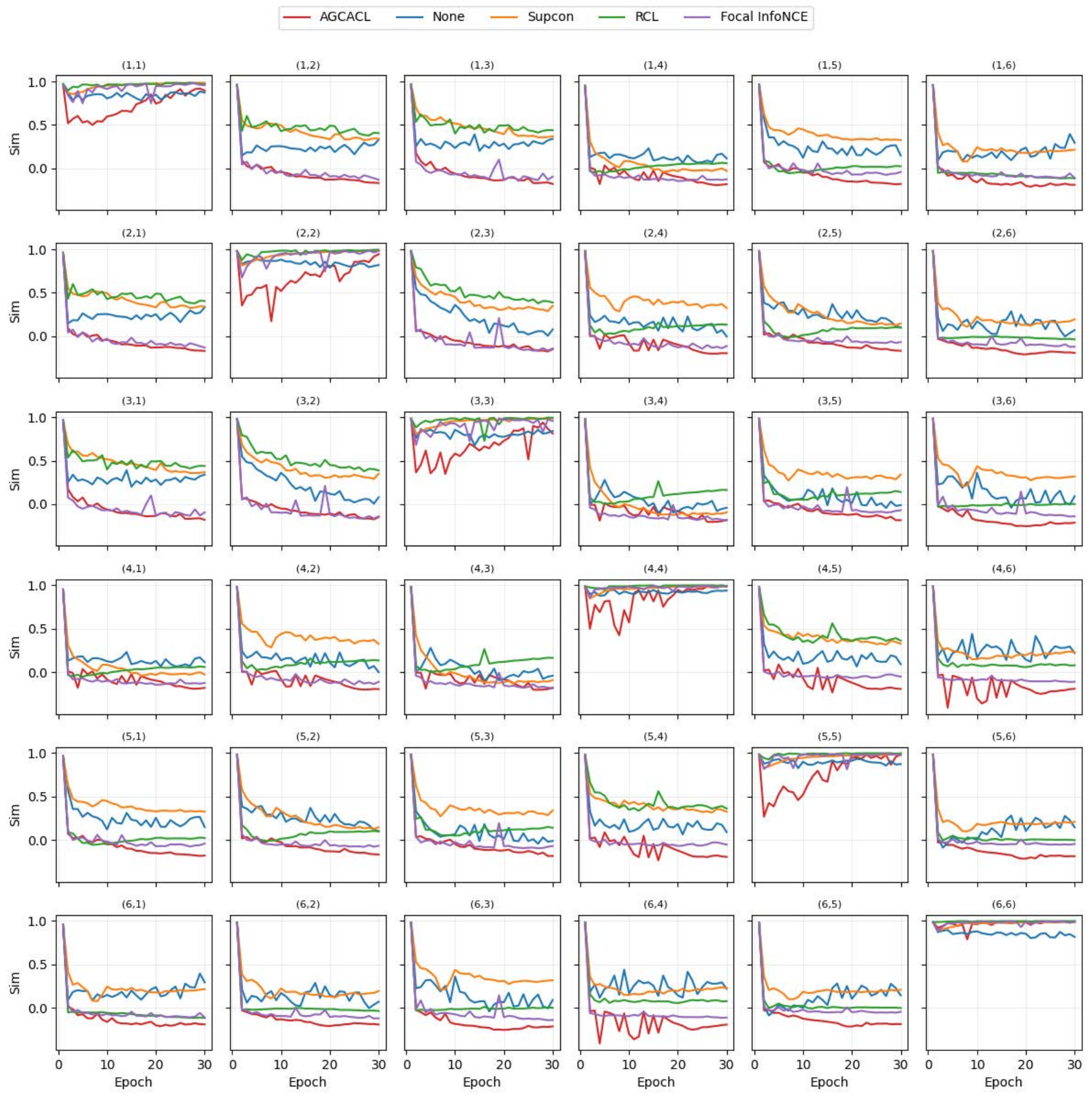}
	\caption{Epoch-wise evolution of pairwise class similarity on the training set for different contrastive objectives. Each subplot corresponds to one class pair; diagonal panels indicate intra-class similarity and off-diagonal panels indicate inter-class similarity. Classes 1–6 are ordered from the head class (sinus rhythm) to the tail class (tachycardias).}
	\label{fig:S_all_epochs}
\end{figure}

\begin{figure}[htbp]
	\centering
	\includegraphics[width=0.95\textwidth]{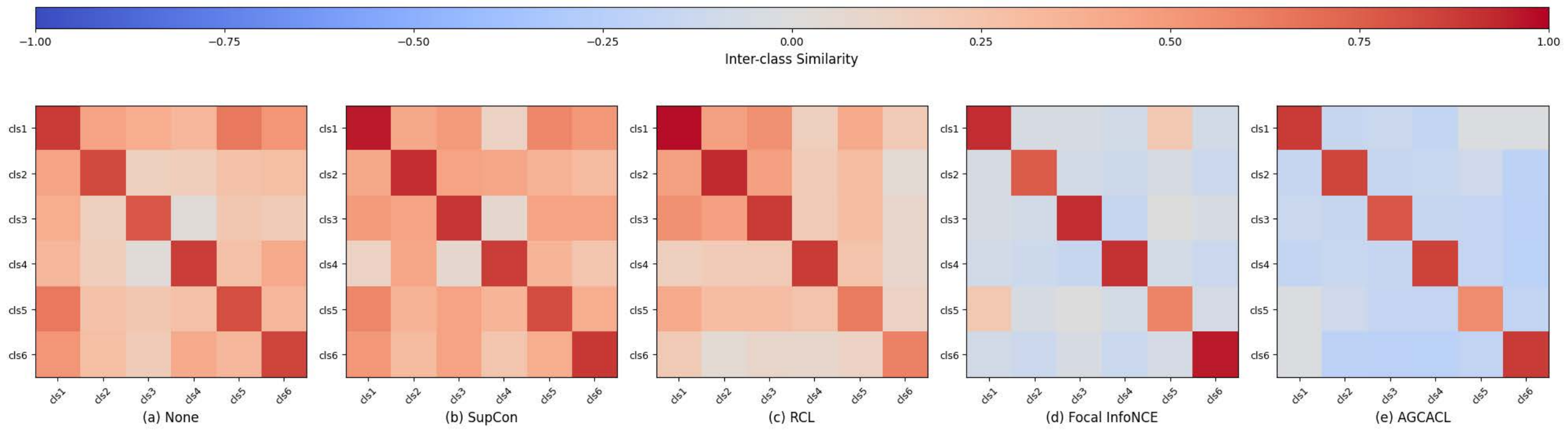}
	\caption{Test-set class-similarity matrices obtained under different contrastive learning strategies. Classes 1–6 are ordered from the head class (sinus rhythm) to the tail class (tachycardias). Lower off-diagonal values indicate stronger inter-class separation.}
	\label{fig:heatmaps}
\end{figure}

\par To further verify the effectiveness of the proposed Adaptive Global Class-Aware Contrastive Loss (AGCACL), we reproduced several representative contrastive learning objectives under a unified experimental framework and compared them with our method. Specifically, we considered four settings: no contrastive loss (None), supervised contrastive loss (SupCon)~\cite{khosla2020supervised}, rebalanced contrastive learning (RCL)~\cite{de2023long}, and Focal InfoNCE~\cite{hou2023improving}. All methods were evaluated using the same backbone, training protocol, data split, and optimization settings, such that the observed differences can be primarily attributed to the loss formulation itself. The purpose of this comparison is to examine how different contrastive objectives influence feature discrimination, especially under the long-tailed and minority-sensitive pediatric arrhythmia setting.

\par As reported in Table~\ref{tab:loss_comparison}, AGCACL achieves the most balanced overall classification performance among all compared loss functions. Although the model without a contrastive term attains a slightly higher Top-1 Accuracy (96.34\% vs.\ 96.22\%), AGCACL yields the best Macro Specificity (98.82\%), Macro Recall (92.57\%), Macro F1 (91.24\%), and Macro F2 (92.01\%), while also improving Macro Precision to 90.16\%. Compared with the no-contrastive baseline, AGCACL improves Macro Precision, Recall, F1, and F2 by 2.06, 4.30, 3.70, and 4.18 percentage points, respectively. This pattern is important because the macro-averaged metrics are more sensitive to class imbalance than Top-1 Accuracy. Therefore, the results indicate that AGCACL does not merely preserve strong overall classification performance, but more importantly enhances the recognition quality of underrepresented and difficult classes. By contrast, RCL achieves the highest Macro Precision (93.04\%) but suffers from a clear drop in Macro Recall (82.64\%), suggesting a relatively conservative decision boundary. Focal InfoNCE improves Macro Recall and Macro F2 over the no-contrastive baseline, but its overall balance between precision and recall remains inferior to that of AGCACL.

\par The differences among existing methods further clarify why. SupCon ignores class imbalance and fails to guarantee sufficient separability for minority classes in the embedding space. RCL introduces class-frequency weighting, but its treatment of hard negatives remains insufficient. Focal InfoNCE improves class separation by emphasizing hard negatives, yet it still suffers from batch-level randomness and lacks global consistency. In contrast, AGCACL unifies class prototypes, global similarity matrices, and adaptive weighting mechanisms within a single framework: it prioritizes separation of the most confusing negative pairs while strengthening intra-class compactness for minority categories with weaker internal consistency. This design enables more stable and targeted optimization of class boundaries.

\par The dynamic evolution of class-wise similarity during training, shown in Fig.~\ref{fig:S_all_epochs}, provides further evidence for this interpretation. For off-diagonal entries, which correspond to inter-class similarity, AGCACL generally drives the curves downward more rapidly and maintains the lowest similarity values in late training, especially for several difficult class pairs. This trend is particularly evident for pairs involving class~6, where AGCACL consistently produces the lowest final similarity against the other classes, indicating that the tail class is more effectively separated from the remaining label space. In contrast, SupCon and RCL often show an initial decrease followed by a noticeable rebound in multiple off-diagonal entries; in some cases, their late-epoch similarity approaches, or even exceeds, the no-contrastive baseline. This behavior suggests that these methods are less stable in maintaining inter-class margins throughout training. Focal InfoNCE performs better than the no-contrastive baseline in many off-diagonal trajectories, which is consistent with its hard-negative emphasis, but its suppression of confusing class pairs is still less persistent and less globally consistent than that of AGCACL.

\par A more subtle phenomenon can be observed on the diagonal entries of Fig.~\ref{fig:S_all_epochs}, which reflect intra-class self-similarity. Compared with the other loss functions, AGCACL exhibits more pronounced fluctuations and, for several classes, an evident drop in the early stage of training. This behavior is especially visible for classes~1--5, where the self-similarity under AGCACL initially falls below the smoother trajectories produced by SupCon, RCL, Focal InfoNCE, and even the no-contrastive baseline. However, this transient decline should not be interpreted as a failure of intra-class compactness. Rather, when considered together with the simultaneous decrease in off-diagonal similarity, it suggests that AGCACL first performs a more aggressive restructuring of the embedding space to open class boundaries, and then progressively restores within-class cohesion as training proceeds. By the end of training, the diagonal similarities under AGCACL recover to a high level across all classes, indicating that the temporary oscillation is the cost of a stronger and more targeted boundary-shaping process, rather than a sign of unstable representation learning.

\par The test-set heatmaps in Fig.~\ref{fig:heatmaps} further confirm the trade-off between inter-class separation and intra-class compactness. The heatmaps of None, SupCon, and RCL still contain many warm off-diagonal regions, indicating that substantial residual similarity remains between different classes. Focal InfoNCE reduces these off-diagonal similarities more effectively, producing a visibly cooler matrix. Nevertheless, AGCACL yields the coldest overall off-diagonal structure while preserving clear diagonal dominance, demonstrating a better balance between class separation and class cohesion. In particular, several of the most confusing class pairs are assigned the lowest similarity under AGCACL, and pairs involving the minority class~6 are especially well separated from the head and medium-frequency classes. This result is consistent with the training-phase trajectories in Fig.~\ref{fig:S_all_epochs} and explains why AGCACL achieves the best macro-level performance in Table~\ref{tab:loss_comparison}.

\par Overall, the comparative results show that AGCACL is not simply another contrastive regularizer, but a loss specifically suited to long-tailed pediatric arrhythmia classification. By combining prototype-based alignment, globally informed hard-class mining, and adaptive class-aware weighting, it improves both the stability and the selectivity of contrastive optimization. As a result, the learned embedding space becomes more compact within classes and more separable across confusing categories, with the clearest benefit appearing on minority and clinically important rhythms.

\subsection{Ablation Studies}
\begin{table}[htbp]
	\centering
	\caption{Ablation results for different modality, fusion, classifier-head, and contrastive-loss configurations. Results are reported using Top-1 accuracy and macro-averaged specificity, precision, recall, F1, and F2.}
	\renewcommand{\arraystretch}{1.3}
	\begin{tabular}{cccc|cccccc}
		\bottomrule
		D & A & T & G & Top-1 Acc & Macro S & Macro P & Macro R & Macro F1 & Macro F2 \\
		\midrule
		&  &  &  & 93.39 & 98.34 & 87.17 & 77.63 & 81.12 & 78.77 \\
		&  &  & $\checkmark$ & 95.49 & 98.73 & 83.91 & 82.88 & 83.29 & 83.02 \\
		&  & $\checkmark$ &  & 95.27 & 98.54 & 77.83 & 74.61 & 75.90 & 75.07 \\
		&  & $\checkmark$ & $\checkmark$ & 94.02 & 98.43 & 79.33 & 86.48 & 82.34 & 84.66 \\
		$\checkmark$ &  &  &  & 95.76 & 98.78 & 82.71 & $\mathbf{93.51}$ & 86.62 & 90.18 \\
		$\checkmark$ &  &  & $\checkmark$ & $\mathbf{96.77}$ & 98.84 & $\mathbf{93.65}$ & 85.52 & 88.95 & 86.78 \\
		$\checkmark$ &  & $\checkmark$ &  & 96.40 & $\mathbf{98.87}$ & 82.78 & 91.36 & 85.12 & 88.01 \\
		$\checkmark$ &  & $\checkmark$ & $\checkmark$ & 95.66 & 98.73 & 84.21 & 91.27 & 86.72 & 89.05 \\
		$\checkmark$ & $\checkmark$ &  &  & 95.19 & 98.65 & 80.93 & 87.46 & 83.15 & 85.34 \\
		$\checkmark$ & $\checkmark$ &  & $\checkmark$ & 96.30 & 98.86 & 82.71 & 90.86 & 85.18 & 87.98 \\
		$\checkmark$ & $\checkmark$ & $\checkmark$ &  & 96.34 & 98.58 & 88.10 & 88.27 & 87.54 & 87.83 \\
		$\checkmark$ & $\checkmark$ & $\checkmark$ & $\checkmark$ & 96.22 & 98.82 & 90.16 & 92.57 & $\mathbf{91.24}$ & $\mathbf{92.01}$ \\
		\bottomrule
	\end{tabular}
	
	\vspace{2mm}
	\footnotesize{\textit{Note:} D $=$ Dual (ECG+IEGM), A $=$ attention fusion in our proposed model, T $=$ Transformer head in our proposed model and G $=$ AGCACL. Checkmark indicates the module is enabled.}
	\label{tab:ablation_study}
\end{table}

\begin{figure}[htbp]
	\centering
	\includegraphics[width=0.95\textwidth]{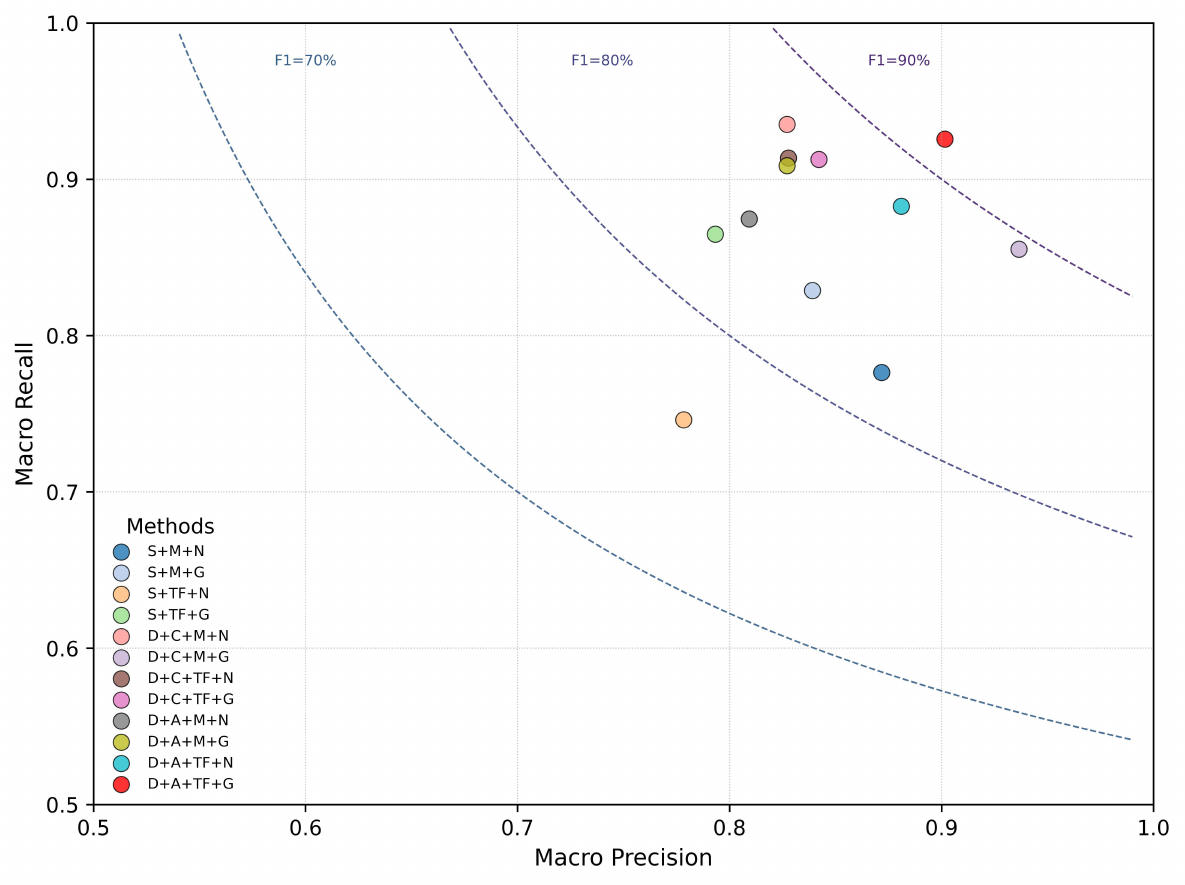}
	\caption{Macro precision–macro recall trade-off across different ablation settings. Each marker denotes one configuration: S, single-modality ECG; D, dual-modality ECG+IEGM; C, concatenation fusion; A, attention fusion; M, MLP head; T, Transformer head; N, no contrastive loss; G, AGCACL.}
	\label{fig:pr}
\end{figure}

\begin{figure}[htbp]
	\centering
	\includegraphics[width=0.95\textwidth]{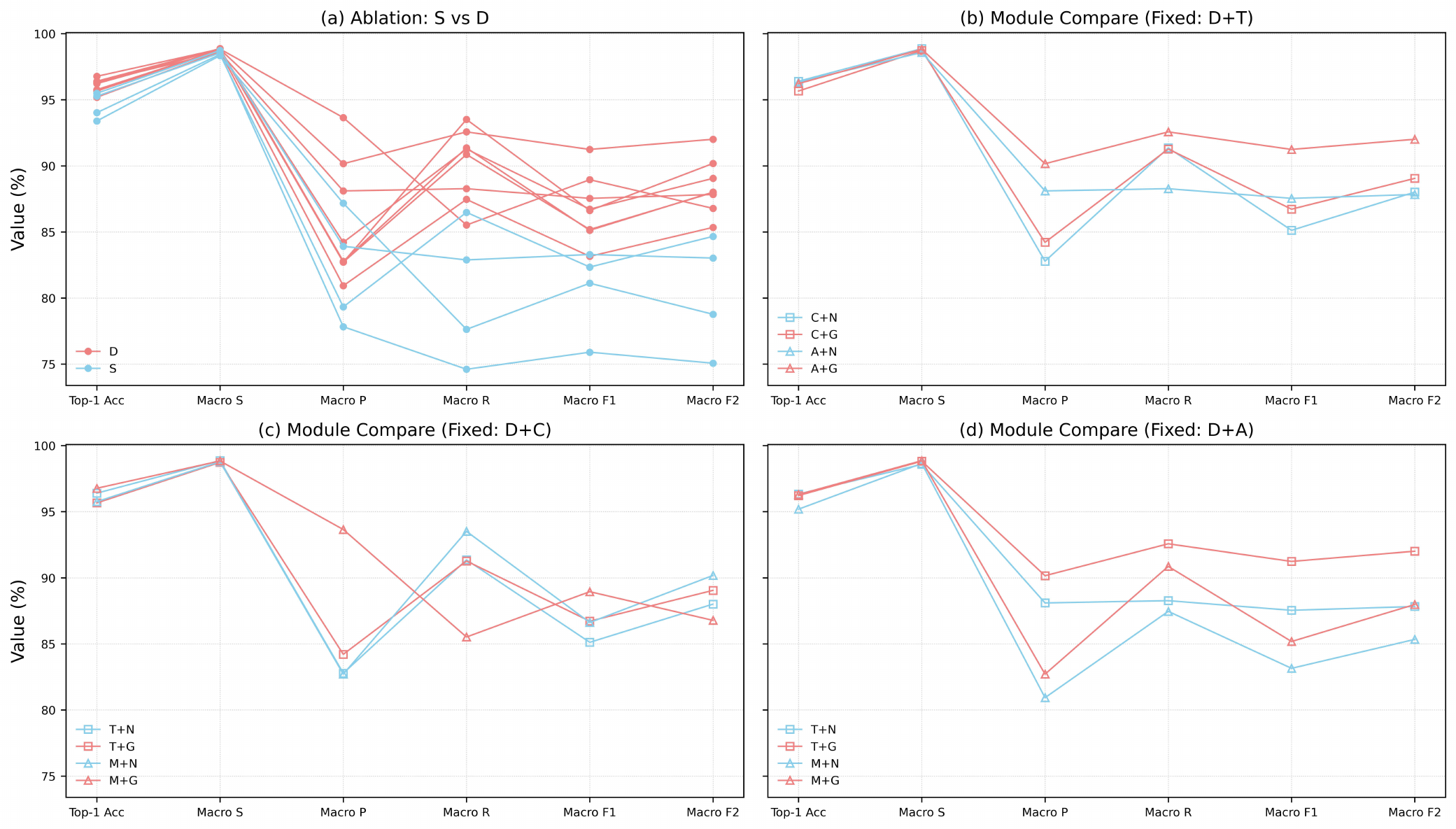}
	\caption{Performance comparison of ablation configurations across six evaluation metrics. Each marker denotes one configuration: S, single-modality ECG; D, dual-modality ECG+IEGM; C, concatenation fusion; A, attention fusion; M, MLP head; T, Transformer head; N, no contrastive loss; G, AGCACL.}
	\label{fig:line}
\end{figure}

\par To clarify the contribution of each design choice in the proposed framework, we conducted a systematic ablation study over four key factors: modality setting, feature fusion strategy, classification head, and contrastive objective. Specifically, we compared Single (surface ECG only) versus Dual (ECG+IEGM), concatenation fusion versus attention-based fusion, MLP versus Transformer head, and no contrastive loss versus AGCACL. The quantitative results are summarized in Table~\ref{tab:ablation_study}, while Fig.~\ref{fig:pr} and Fig.~\ref{fig:line} visualize the precision--recall trade-offs and module interaction patterns under different configurations.

\par First, the benefit of multimodality is clear and consistent. Across matched settings, Dual models generally outperform their Single counterparts on the macro-averaged metrics, especially Macro Recall, Macro F1, and Macro F2. For example, under the MLP head without contrastive learning, moving from Single to Dual improves Top-1 Accuracy from 93.39\% to 95.76\%, Macro Recall from 77.63\% to 93.51\%, Macro F1 from 81.12\% to 86.62\%, and Macro F2 from 78.77\% to 90.18\%. A similar trend is observed under the Transformer head without contrastive learning, where the Dual configuration improves Macro Recall by 16.75 percentage points and Macro F1 by 9.22 percentage points over the corresponding Single model. These results indicate that IEGM provides complementary electrophysiological information beyond surface ECG alone, and that multimodal learning is particularly beneficial for recognizing difficult and minority rhythm categories.

\par Second, AGCACL improves the macro-level performance in a highly consistent manner. As shown in Table~\ref{tab:ablation_study}, replacing the standard classification objective with AGCACL increases Macro F1 in all six matched comparisons and increases Macro F2 in five of the six settings. The gains are especially notable in configurations with stronger representational capacity, such as Single+Transformer and Dual+Attention+Transformer. For instance, under the Single+Concatenation+Transformer setting, AGCACL improves Macro Recall from 74.61\% to 86.48\%, Macro F1 from 75.90\% to 82.34\%, and Macro F2 from 75.07\% to 84.66\%. Under the final Dual+Attention+Transformer setting, AGCACL further improves Macro Precision, Recall, F1, and F2 by 2.06, 4.30, 3.70, and 4.18 percentage points, respectively. As also illustrated in Fig.~\ref{fig:pr}, configurations equipped with AGCACL generally move upward and to the right in the precision--recall plane, indicating a more favorable balance between sensitivity and reliability. Although AGCACL does not monotonically improve every single metric in every architecture, its effect on recall-sensitive and imbalance-sensitive metrics is clearly more robust than that on Top-1 Accuracy alone.

\par Third, the interaction between attention fusion and the Transformer head reveals an important synergy. Attention fusion by itself does not universally dominate simple concatenation; when followed by an MLP head, its gains are limited and can even be inconsistent. This suggests that bridging the semantic gap between ECG and IEGM is only part of the problem, and that the aligned multimodal representation still requires a sufficiently expressive classifier to exploit the refined cross-modal structure. Once the Transformer head is introduced, however, the advantage of attention fusion becomes much clearer. Under the Dual+Transformer setting with AGCACL, replacing concatenation with attention fusion improves Top-1 Accuracy from 95.66\% to 96.22\%, Macro Precision from 84.21\% to 90.16\%, Macro Recall from 91.27\% to 92.57\%, Macro F1 from 86.72\% to 91.24\%, and Macro F2 from 89.05\% to 92.01\%. This result strongly suggests that attention fusion and the Transformer head are complementary: the former promotes semantically aligned multimodal integration, while the latter captures higher-order global dependencies within the fused representation.

\par A similar conclusion can be drawn when fixing the Dual+Attention setting and comparing the classification heads. Without AGCACL, replacing the MLP with the Transformer increases Macro Precision from 80.93\% to 88.10\% and Macro F1 from 83.15\% to 87.54\%. With AGCACL, the same replacement further improves Macro Precision from 82.71\% to 90.16\%, Macro Recall from 90.86\% to 92.57\%, Macro F1 from 85.18\% to 91.24\%, and Macro F2 from 87.98\% to 92.01\%. Therefore, the Transformer head is not merely an alternative classifier, but a key component that allows the model to better exploit multimodal and contrastively structured features. This observation is also reflected in Fig.~\ref{fig:line}, where the strongest curves consistently emerge when attention fusion, Transformer modeling, and AGCACL are combined.

\par Overall, the ablation results support a coherent causal interpretation of the final architecture. Multimodality supplies richer and more complementary physiological information; attention fusion reduces the semantic discrepancy between ECG and IEGM; the lightweight Transformer head captures global dependencies in the fused representation; and AGCACL regularizes the embedding space toward better minority-class compactness and inter-class separation. Their combination leads to the final configuration (Dual+Attention+Transformer+AGCACL), which achieves the best overall macro-level performance in Table~\ref{tab:ablation_study}, with 96.22\% Top-1 Accuracy, 98.82\% Macro Specificity, 90.16\% Macro Precision, 92.57\% Macro Recall, 91.24\% Macro F1, and 92.01\% Macro F2. Importantly, these gains are obtained while maintaining highly competitive Top-1 Accuracy and Macro Specificity, indicating that the final model improves minority-sensitive recognition without sacrificing overall diagnostic reliability. The full configuration does not maximize every individual metric. In particular, some lighter variants achieve slightly higher Top-1 accuracy or Macro Precision. However, the complete model provides the best overall balance on minority-sensitive metrics, especially Macro Recall, Macro F1, and Macro F2, which are more aligned with the goal of long-tailed pediatric arrhythmia recognition.

\subsection{Limitations and Generalizability}

\par This study has several data-related limitations. First, although the Leipzig Heart Center ECG-Database is highly valuable for pediatric arrhythmia research, the pediatric subset used here contains only 29 subjects. Owing to this limited number of patients and the extreme sparsity of several rhythm categories, the dataset was partitioned at the window level rather than by strictly subject-independent splitting, so as to preserve class coverage in the training, validation, and test sets. While this design makes the present benchmark feasible under severe long-tailed conditions, it may still introduce residual intra-subject correlation across splits and thus yield more optimistic estimates than a fully patient-level evaluation. In addition, the data come from a single center, which may limit the representativeness of signal patterns, acquisition conditions, and patient heterogeneity.

\par Second, the label distribution of the pediatric subset is extremely imbalanced, with only a few episodes and short cumulative durations for several minority rhythms. This not only increases the difficulty of robust evaluation, but also limits the granularity of the classification task itself, since some low-level rhythm labels are absent or too rare to be modeled reliably and therefore had to be merged into high-level categories. As a result, although the present study provides a practical and reproducible benchmark for long-tailed pediatric arrhythmia classification, larger multicenter pediatric datasets with more subjects, richer minority-class coverage, and subject-independent splits are still needed to more rigorously assess generalization and clinical utility.

\section{Conclusion}\label{Conclusion}

\par In this study, we developed a multimodal deep learning framework for long-tailed pediatric arrhythmia classification on the Leipzig Heart Center ECG-Database. To address the challenges of limited pediatric data, severe class imbalance, and high confusion among minority rhythms, we established a reproducible preprocessing pipeline and proposed a dual-branch architecture that integrates ECG and IEGM through ResNet-based encoders, semantic attention fusion, and a lightweight Transformer head. We further introduced the Adaptive Global Class-Aware Contrastive Loss (AGCACL) to enhance intra-class compactness and inter-class separability under long-tailed conditions.

\par Experimental results showed that the proposed method achieved state-of-the-art performance on this dataset, with 96.22\% Top-1 Accuracy, 98.82\% Macro Specificity, 90.16\% Macro Precision, 92.57\% Macro Recall, 91.24\% Macro F1, and 92.01\% Macro F2. The gains were especially evident on macro-level metrics, indicating improved recognition of underrepresented and clinically important rhythms. Overall, these findings suggest that combining multimodal learning with class-aware contrastive optimization is a promising direction for pediatric arrhythmia analysis. Future work will focus on validation with larger pediatric cohorts, subject-independent evaluation, and broader assessment of clinical generalizability.

\section*{Acknowledgment}
The authors would like to thank the team who made the dataset publicly available, as well as the patients, their families, and the clinical staff of the affiliated hospital for their contributions and support. This work was supported by the Basic Research Fund in Shenzhen Natural Science Foundation (No.\ JCYJ20240813104924033).

\section*{Author contributions}
\textbf{Yiqiao Chen}: 
Conceptualization,
Project administration,
Methodology,
Formal analysis,
Investigation,
Data curation,
Writing – Original Draft,
Visualization.
\textbf{Zijian Huang}: 
Conceptualization,
Project administration,
Methodology,
Formal analysis,
Investigation,
Data curation,
Writing – Original Draft,
Visualization.
\textbf{Zhenghui
Feng}: 
Supervision,
Project administration,
Funding acquisition.

\section*{Conflict of Interest}
The authors declare that they have no conflict of interest.

\section*{Data Availability}
The dataset analyzed in this study is publicly available from the Leipzig Heart Center ECG-Database via PhysioNet \cite{klehsleipzig, goldberger2000physiobank}:
\url{https://www.physionet.org/content/leipzig-heart-center-ecg/1.0.0/}

\bibliographystyle{elsarticle-num}

\end{document}